\documentclass[aps,print,showpacs]{revtex4}%
\usepackage{amsfonts}
\usepackage{amsmath}
\usepackage{amssymb}
\usepackage{graphicx}%
\providecommand{\U}[1]{\protect\rule{.1in}{.1in}}

\begin{document}
\title{Quantum droplets in two-dimensional optical lattices}
\author{Yiyin Zheng$^{\text{a}}$, Shantong Chen$^{\text{a}}$, Zhipeng Huang, Shixuan
Dai, Bin Liu, Yongyao Li}
\email{autumnlyy@126.com}
\author{Shurong Wang}
\affiliation{School of Physics and Optoelectronic Engineering, Foshan University, Foshan
528000, China }
\affiliation{$^{\text{a}}$ These authors contributed equally to this work.}


\begin{abstract}
We study the stability of zero-vorticity and vortex lattice quantum droplets
(LQDs), which are described by a two-dimensional (2D) Gross-Pitaevskii (GP)
equation with a periodic potential and Lee-Huang-Yang (LHY) term. The LQDs are
divided in two types: onsite-centered and offsite-centered LQDs, the centers
of which are located at the minimum and the maximum of the potential,
respectively. The stability areas of these two types of LQDs with different
number of sites for zero-vorticity and vorticity with $S=1$ are given. We
found that the $\mu-N$ relationship of the stable LQDs with a fixed number of
sites can violate the Vakhitov-Kolokolov (VK) criterion, which is a necessary
stability condition for nonlinear modes with an attractive interaction.
Moreover, the $\mu-N$ relationship shows that two types of vortex LQDs with
the same number of sites are degenerated, while the zero-vorticity LQDs are
not degenerated. It is worth mentioning that the offsite-centered LQDs with
zero-vorticity and vortex LQDs with $S=1$ are heterogeneous.


\end{abstract}
\maketitle

\section{Introduction}

In Bose-Einstein condensates (BECs), it is well known that free-space
nonlinear modes may collapse in two-dimensional ($2$D) and three-dimensional
($3$D) geometries via the action of the usual attractive cubic nonlinearity
\cite{quantun collapse}. Hence, how to acquire stabilize nonlinear modes in
multidimensional systems remains an important research topic. Generally, the
simplest way is to modify the attractive cubic nonlinearity, which includes
reducing the cubic nonlinearity to quadratic nonlinearity
\cite{Torruellas,Xliu2000}, adding competitive nonlinearities, such as the
competing cubic-quintic nonlinearity
\cite{Mihalache2001,CQ2,ELF2013,WYYDN90,DCQND87,ND93_2379,ND91_757}, changing
cubic nonlinearity to saturable nonlinearity \cite{Segev1994} or nonlocal
nonlinearity
\cite{Peccianti2002,Pedri2005,Tikonenkow2008,Jiasheng,Maucher2011} etc.

Subsequently, one can introduce spin-orbit coupling to stabilize self-trapped
modes, i.e., matter-wave solitons
\cite{Xuyong2015,Xunda2016,Yongyao20172,ND_CXW,cfs_yzj,njp_lb,Bingjin2017,Bingjin2018,Shimei2018,Weipang2018}
and quantum droplets (QDs) \cite{PRA98_023630}. QDs, a new type of
self-bound quantum liquid state, were created experimentally in dipolar
bosonic gases of dysprosium \cite{Schmitt2016} and erbium \cite{Chomaz2016},
as well as in mixtures of two atomic states of $^{39}$K \cite{Cabrera2018}
with contact interactions. QDs have attracted much attention in the field of
ultracold atoms
\cite{PetrovPRL115_155302,PetrovPRL117_100401,PRA98_051603R,PRA98_051604,2DvortexQD,PRL_LYY,LB_CNSNS,
LB_PRA,Zhouzheng_cnsns,Zhouzheng_cnsns2,Wachtler2016,Wachtler20162,Daillie2016,PRL119_050403,PRL116_215301,PRL120_160402,unstable-vort-DD,PRA94_033619,PRA97_011602,Staudinger2018,Cikojevic2018,Astrakharchik2018,Cabrera2,Inguscio,PRA97_053623,PRA97_063616,Luca-Rabi,PRA98_053835,Boris_Condens.
Matter,LZH,SOCQD}, which predicate a possibility in the framework of the 3D
\cite{PetrovPRL115_155302}, 2D
\cite{PetrovPRL117_100401,PRA98_051603R,PRA98_051604,2DvortexQD,SOCQD,PRL_LYY,LB_CNSNS}
and 1D \cite{LB_PRA,Zhouzheng_cnsns,Zhouzheng_cnsns2,Astrakharchik2018}
Gross-Pitaevskii (GP) equations. Research has shown that QDs have been formed
with the help of zero-point quantum fluctuations, which can arrest the
collapse of attractive Bose gases in two and three dimensions and can be
described theoretically by the Lee-Huang-Yang (LHY) correction \cite{LHY}. The
LHY correction is proportional to $n^{5/2}$ in $3$D \cite{PetrovPRL115_155302}
and $n^{2}\ln\left(  n/\sqrt{e}\right)  $\ in $2$D \cite{PetrovPRL117_100401},
where $n$ is the condensate density, which plays an important role in
stabilizing QDs. In these studies, in the $2$D and $3$D domains, the LHY term
acts as a higher-order nonlinear repulsive interaction in the relative GP
equations, which arrest the collapse of attractive Bose gases induced by the
mean-field force.

Another possibility for the stabilization of nonlinear modes was revealed in
BECs, which are trapped in an optical lattice \cite{BEC_OL}. Numerous studies
have shown that BECs trapped in an optical lattice provide an ideal and clean
platform for studying the nonlinear modes and their dynamics. In the optical
lattice, various types of matter-wave solitons were investigated both
numerically and analytically
\cite{ZHF_CPB,PRA96_043626,ZX_JPB,CPL30_060306,AML92_15,ActaPS63_190502,annals
of
p322_1961,CPB21_020306,PRA71_053611,pra99_023630,ActaPS68_043703,PRA95_043618,mplb32_1850070,PRA98_033827,cpl33_110301}%
. A potentially interesting research direction is to study QDs in a lattice.
Recently, dynamics of QDs in a 1D optical lattice were studied
\cite{Zhouzheng_cnsns}, and it was found that the optical lattice potential
strongly influences the stability of the QDs, and this system can support
stable dipole-model QDs in a very small range of proper parameters. In the 2D
or 3D domains, optical lattices can provide more freedom than their 1D
counterparts. For example, in higher-dimensional spaces, we can construct more
complex lattice structures, such as square lattices
\cite{squarelattice1,squarelattice2,squarelattice3,squarelattice4,squarelattice5}%
, triangle lattices \cite{trianglelattice1,trianglelattice2,trianglelattice3}
and honeycomb lattices
\cite{honeycomblattice1,honeycomblattice2,honeycomblattice3,honeycomblattice4}
(graphene structure). Furthermore, in the higher-dimensional domains, we can
not only consider the fundamental or the dipole modes but also the quadrupole
modes \cite{quadrupole1,quadrupole2,quadrupole3,quadrupole4,quadrupole5}, or
more interestingly, the vortex modes. However, the dynamics of QDs in these
higher-dimensional spaces with periodic potential have not been considered
thus far. It was known that periodic potential is a fundamental problem in
solid-state physics. The combination of QDs and lattice potential may open an
avenue to study the dynamics of such new kind of liquid in some advanced topic
of condense matter physics, such as, discrete systems, topological objects, etc.

The objective of the present work is to demonstrate the possibility of
creating stable zero-vorticity and vortex LQDs in BECs, which are trapped in a
$2$D optical square lattice, and study their characteristics. Similar to their
1D counterparts, two types of LQDs, viz., onsite-centered and offsite-centered
LQDs, the centers of which are located at the minimum and the maximum of the
potential, respectively, are identified. The stability of these LQDs with
zero-vorticity ($S=0$) and a vortex with $S=1$ are studied in detail. The rest
of this paper is structured as follows: the model for the current system is
described in Section II, and the results of the 2D LQDs for $S=0$ and $1$ are
discussed in detail in Section III, and this work is concluded in Section IV.

\section{The model}

We assume that the LQDs, which are formed by binary BECs, are strongly
confined in the transverse direction with lateral size $l\gg\sqrt{a_{\pm
}a_{\perp}}$, where $a_{\pm}$ and $a_{\perp}$ are the self-repulsion
scattering lengths of each component and the transverse confinement length,
respectively. For the current experimental system, we can select the length of
a single lattice site to be $D\sim1$ $\mu$m, $a_{\pm}\sim3$ nm, and $a_{\perp
}\ll1$ $\mu$m. This condition definitely holds for $l\sim10D$. In this case,
the GP equation with LHY correction is also reduced to 2D form for the scaled
wave functions $\psi_{\pm}$ of the two components as
\begin{equation}
i\partial_{t}\psi_{\pm}=-{\frac{1}{2}}\nabla^{2}\psi_{\pm}+{\frac{4\pi}{g}%
}(|\psi_{\pm}|^{2}-|\psi_{\mp}|^{2})\psi_{\pm}+V(x,y)\psi_{\pm}+(|\psi_{\pm
}|^{2}+|\psi_{\mp}|^{2})\psi_{\pm}\ln(|\psi_{\pm}|^{2}+|\psi_{\mp}|^{2}),
\label{Full-GPE}%
\end{equation}
where $g>0$ is the coupling constant and $V(x,y)$ is the 2D lattice potential.
Here, the logarithmic form of the LHY term has been adopted because of strong
transverse confinement. If these two components are under a symmetric
condition, i.e., $a_{+}=a_{-}$, the mean-field self-repulsion and the
cross-attraction between the binary BECs are canceled with each other, Eq.
(\ref{Full-GPE}) can be further simplified by adopting the symmetric state
$\psi_{+}=\psi_{-}=\phi/\sqrt{2}$ as:
\begin{equation}
i\frac{\partial\phi}{\partial{t}}=-{\frac{1}{2}}{{\nabla}^{2}}\phi
+V(x,y)\phi+|\phi|^{2}\text{ln}|\phi|^{2}\phi. \label{Model}%
\end{equation}
The analysis in Ref. \cite{2DvortexQD} has demonstrated that the energy of
this LHY term can provide a minimum for supporting a QDs. Adding the lattice
potential does not eliminate such an energy minimum, hence, the LQDs can
formed in the current model.

Here, we use the square lattices as the potential; hence, $V(x,y)$ can be
described as:%
\begin{equation}
V(x,y)=V_{0}\left[  \cos^{2}\left(  {\frac{\pi}{D}}x\right)  +\cos^{2}\left(
{\frac{\pi}{D}}y\right)  \right]  \label{v_cos}%
\end{equation}
or
\begin{equation}
V(x,y)=V_{0}\left[  \sin^{2}\left(  {\frac{\pi}{D}}x\right)  +\sin^{2}\left(
{\frac{\pi}{D}}y\right)  \right]  \label{v_sin}%
\end{equation}
where $V_{0}<0$ is the lattice modulation depth and $D$ is the lattice
constant (i.e., the period of the lattice). Refer to Ref. \cite{book_BEC in
dilute gases}, the simplest way to form 1D lattices is by superimposing two
oppositely directed laser beams with the same frequency, and may be produced
in higher-dimensions by superimposing more than two beams with different wave
vectors. We consider the case in which the polarizations of the beams
propagating in the $x$-direction are the same, as are the polarizations of the
beams propagating in the $y$-direction, but the polarization of the beams
propagating in the two directions are orthogonal. The resulting potential
energy becomes proportional to $\cos^{2}(qx)+\cos^{2}(qy)$ or $\sin
^{2}(qx)+\sin^{2}(qy)$ and gives rise to a square lattice with lattice
constant equal to $D=\lambda/2$. Even though there is no essential difference
in the expressions in Eqs. (\ref{v_cos}) and (\ref{v_sin}) for square
lattices, which feature only a $\pi/2$ phase shift, the two expressions in
Eqs. (\ref{v_cos}) and (\ref{v_sin}) are crucial for the formation of the LQDs
centered at the origin of the coordinates. Because we let $V_{0}<0$, the
square lattice constructed by Eq. (\ref{v_cos}) at the origin of the
coordinates is a local minimum of the potential, while the lattice
constructing by Eq. (\ref{v_sin}) at the origin of the coordinates is a local
maximum of the potential, which means that the square lattice constructed by
Eq. (\ref{v_cos}) can provide a lattice site for the center of the droplets,
while the lattice of Eq. (\ref{v_sin}) cannot. Therefore, we can use these two
expressions, Eqs. (\ref{v_cos}) and (\ref{v_sin}), to characterize the two
types of LQDs, which are centered at the minimum and the maximum of the
potential, respectively. The former (the center located at the minimum of the
potential) are called onsite-centered LQDs, and the latter (the center located
at the maximum of the potential) are called offsite-centered LQDs. The total
norm for these two types of LQDs can all be characterized as
\begin{equation}
N={\int{\int}}\left\vert \phi\right\vert ^{2}{dxdy.} \label{Norn}%
\end{equation}

Stationary solutions to Eq. (\ref{Model}) with a chemical potential $\mu$ are
sought as follows:%
\begin{equation}
\phi\left(  {x,y,t}\right)  =\tilde{\phi}\left(  {x,y}\right)  {e^{-i\mu t},}
\label{QDs_solution}%
\end{equation}
where $\tilde{\phi}\left(  {x,y}\right)  $ represents the stationary wave
function. Stationary LQDs with different topological charges $S$ were produced
by the imaginary-time-integration (ITM) method \cite{ITP1,ITP2} with an
initial guess of:%
\begin{equation}
\phi_{0}\left(  {x,y,t}\right)  =CR^{S}\exp\left(  iS\theta-\alpha
R^{2}\right)  , \label{initial}%
\end{equation}
where $C$ and $\alpha$ are positive real numbers, $R$ and $\theta$ are 2D
polar coordinates. Then, the stability of the solutions was verified by direct
simulations of Eq. (\ref{Model}) in real time, with 1\% random noise is added
in the amplitude given by solutions. The soliton is stable if its density
profile remains unchanged throughout the simulations. Here, we provide some
analysis of the variation of the size of LQDs depends on $N$. Assuming the
LQDs have occupied many lattices, we can use $V_{0}/2$ to average the lattice
potentials in Eqs. (\ref{v_cos}) and (\ref{v_sin}), hence, the GP equation in
Eq. (\ref{Model}) becomes:
\begin{equation}
i\frac{\partial\phi}{\partial{t}}=-{\frac{1}{2}}{{\nabla}^{2}}\phi+{\frac
{1}{2}}V_{0}\phi+|\phi|^{2}\text{ln}|\phi|^{2}\phi. \label{Model2}%
\end{equation}
The energy of the system is:
\begin{equation}
E={\frac{1}{2}}\int\int\left[  |{\nabla}\phi|^{2}+V_{0}|\phi|^{2}+|\phi
|^{4}\text{ln}\left(  {\frac{|\phi|^{2}}{\sqrt{e}}}\right)  \right]  dxdy.
\label{Energy}%
\end{equation}
According to Ref. \cite{2DvortexQD}, Eqs. (\ref{Model2}) and (\ref{Energy})
can give rise to flat-top LQDs if the Thomas-Fermi (TF) approximation is
adopted, and a constant peak density $|\phi_{p}|^{2}$ can be produced. Hence,
the total area of the LQDs can be estimated by:
\begin{equation}
\mathcal{A}={\frac{N}{|\phi_{p}|^{2}}} \label{totalarea}%
\end{equation}
Assuming $\mathcal{A}\approx ns$, where $s$ is the area per lattice site and
$n$ is the number of the lattices occupied by the LQDs, Eq. (\ref{totalarea})
becomes:
\begin{equation}
n={\frac{N}{s|\phi_{p}|^{2}}}, \label{n-N}%
\end{equation}
which indicates that the total number of sites in the LQDs increases as $N$ increases.

In the next section, we will study the effect of the optical lattice potential
on the stationary LQDs in detail. For convenience, we fix the value of the
lattice constant $D=5$ in the numerical simulations. Hence, the free control
parameters in the system are lattice modulation depth $V_{0}$ and the total
norm of the LQDs $N$.

\section{Numerical results}

\subsection{Zero-vorticity LQDs}

\begin{figure}[ptb]
\centering
\includegraphics[scale=0.3]{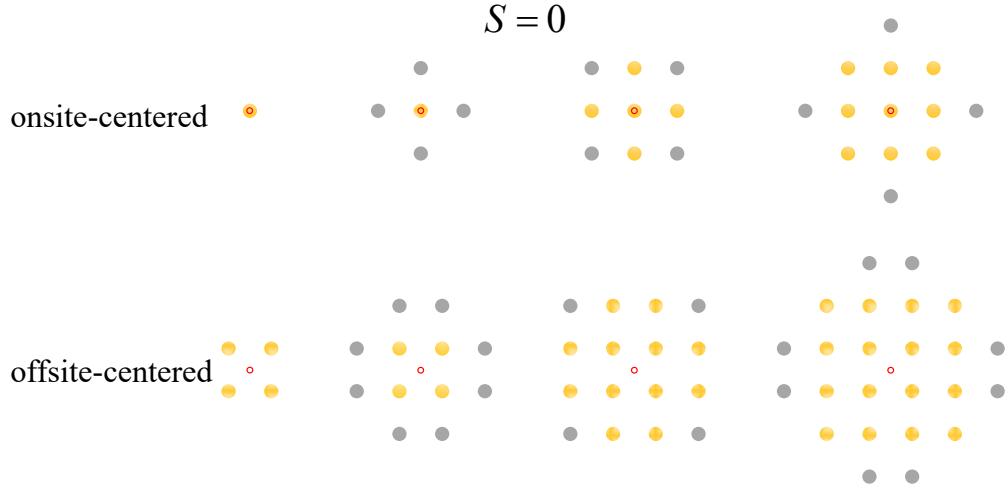}\caption{The upper row is the sketch map
of the growth pattern of onstie-centered LQDs ($S=0)$ versus $n$. The lower
row is the same sketch map for offsite-centered LQDs. The red circles
represent the centers of the LQDs.}%
\label{0}%
\end{figure}

\begin{figure}[ptb]
\centering
\includegraphics[scale=0.8]{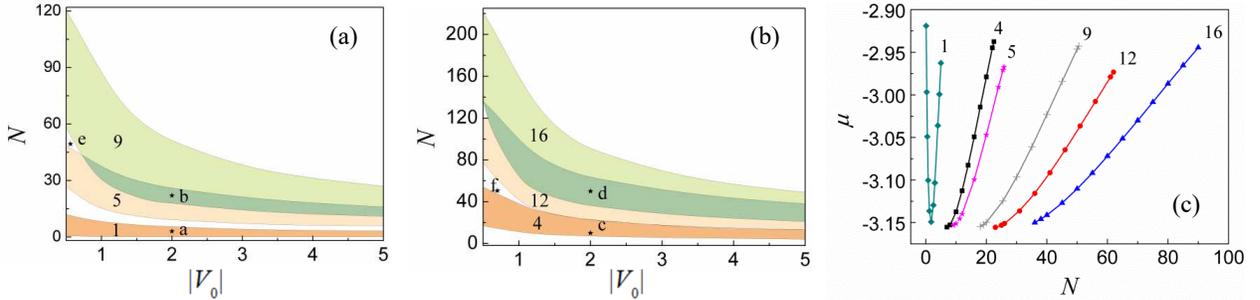}\caption{(a, b) Stability areas of the
minimum three \ LQDs with zero-vorticity, which correspond to $1$, $5$, and
$9$ sites for the onsite-centered LQDs and $4$, $12$, and $16$ sites for the
offsite-centered LQDs in the plane of ($N$, $|V_{0}|$). The bistability states
occur in the light green areas. (c) Chemical potential $\mu$ versus the total
norm $N$, for stable onsite-centered and offsite-centered LQDs. Here, we fixed
$V_{0}=-2$.}%
\label{1}%
\end{figure}Even though Eq. (\ref{n-N}) demonstrates that the total number of
sites of the LQDs increase as $N$ increases for the onsite-centered LQDs and
offsite-centered LQDs, the amounts of growth in $n$ are different. Fig.
\ref{0} shows the sketch maps of the growth pattern of the first 4 steps for
these two types of LQDs. For the onsite-centered LQDs, the $1^{\mathrm{st}}$
step of its pattern occupies one lattice site (i.e., $n=1$) at the origin of
the coordinates. The next step is the extension from the center to the two
sides in the $x$- and $y$- directions ($n=5$). In the $3^{\mathrm{rd}}$ step,
it occupies the four corners, and it becomes a 9-site square. Then, it
continues to expand to the two sides in $x$- and $y$-directions in the
$4^{\mathrm{th}}$ step and $n=13$. For the offsite-centered LQDs, because the
center are not occupied by a lattice site, the $1^{\mathrm{st}}$ step is a
4-site square shape with four closest lattice sites situated around the origin
of the coordinates. Then, it expands to the two sides in the $x$- and
$y$-directions in the $2^{\mathrm{nd}}$ step, and $n=12$. In the
$3^{\mathrm{rd}}$ step, the offsite-centered LQDs become a 16-site square
shape again by filling the 4 corners. Then, it becomes $n=24$ by keeping the
expansion to the two sides in $x$- and $y$-directions in the $4^{\mathrm{th}}$
step. It is interesting to note that the number of sites for the
zero-vorticity onsite-centered LQDs are odd, while the numbers of sites for
the zero-vorticity offsite-centered LQDs are even. Following the above
description, it is convenient to use the number of sites, $n$, to characterize
the LQDs in the following study.

The stationary solutions of the LQDs with different numbers of sites are
carried out by numerically solving Eq. (\ref{Model}). The stabilities of the
LQDs are identified by the direct simulation of Eq. (\ref{Model}) initialized
with the solution by adding 1\% random noise. Numerical studies found that the
stability areas for the two types of LQDs, which are characterized by the
number of sites, depend on total norm $N$ and the modulation depth of the
potential $V_{0}$. Figs. \ref{1}(a,b) display the stability areas of the
onsite-centered and offsite-centered LQDs, respectively, in steps
$1\rightarrow3$ in the $(N,\left\vert V_{0}\right\vert )$ plane. According to
Fig. \ref{0}, the number of sites for the onsite-centered LQDs are $1$, $5$,
and $9$, while the number of sites for the offsite-centered LQDs are $4$, $12$
and $16$. Figs. \ref{1}(a,b) show that these number as color stripes, which
represent the stability areas of the LQDs for different numbers of sites,
concentrated to the lower $N$ area during the increase of $\left\vert
V_{0}\right\vert $. Those phenomena can be explained as follow: the increase
in the modulation depth decreases the effective area per lattice site (i.e.,
$s$), and according to Eq. (\ref{n-N}), resulting in a corresponding increase
in the growth rate of $n$ versus $N$. Moreover, the concentration of these
color stripes gives rise to the overlap between them [see the light green
areas in Figs. \ref{1}(a,b)]. These areas of overlap allow the coexistence of
the different numbers of sites with equal norm $N$ and lattice modulation
depth $V_{0}$.

\begin{figure}[ptb]
\centering
{\includegraphics[width=0.8\columnwidth=1]{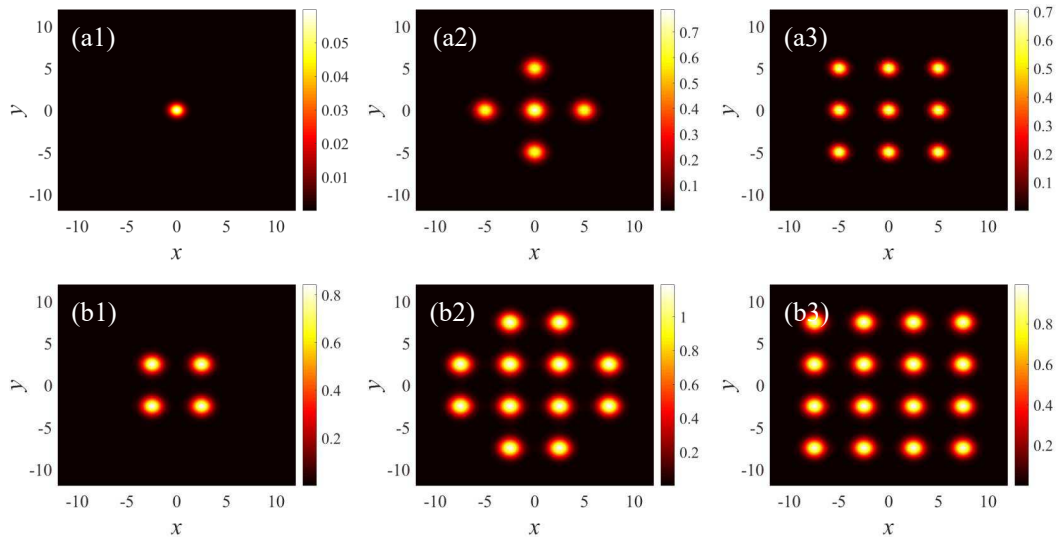}}\caption{(Color online)
Typical examples of the density patterns of the stable onsite-centered (a1-a3)
and offsite-centered (b1-b3) LQDs with zero-vorticity, which correspond to
points \textquotedblleft a, b, c, and d\textquotedblright\ in the stability
areas in Fig. \ref{1}. The parameters are $(N,V_{0})=(3,-2)$, $(22,-2)$,
$(10,-2)$, and $(50,-2)$, respectively. Here, the LQDs in (a2,a3) and (b2,b3)
are selected from the bistable areas [see the points \textquotedblleft
b\textquotedblright\ and \textquotedblleft d\textquotedblright\ in Figs.
\ref{1}(a,b) respectively].}%
\label{2}%
\end{figure}\begin{figure}[ptb]
\centering
{\includegraphics[width=0.6\columnwidth=1]{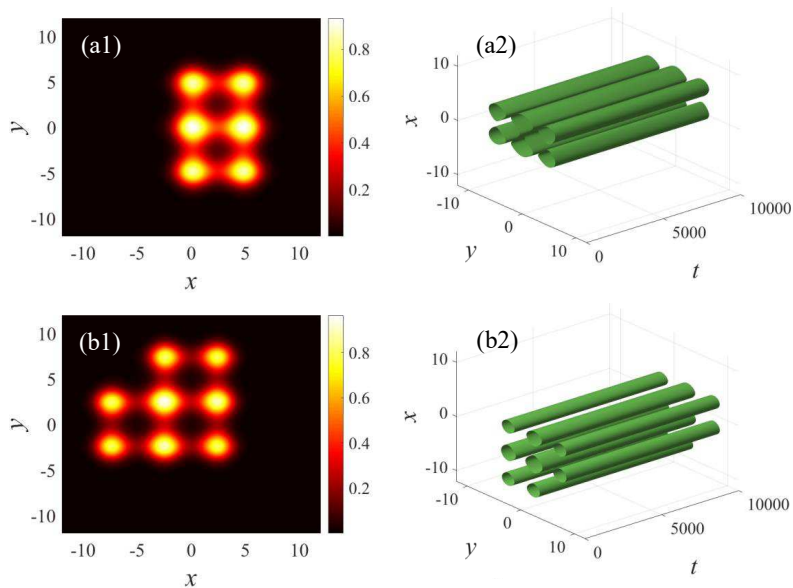}}\caption{Typical examples
of some other LQDs solutions in the white region, which are not symmetric to
the origin of the coordinates, whose parameters are $(N,V_{0})=(49.4,-0.56)$
in (a1 and a2) and $(50.5,-0.7)$ in (b1 and b2), which correspond to points
\textquotedblleft e and f\textquotedblright\ in the blank areas in Figs.
\ref{1}(a, b).}%
\label{3}%
\end{figure}

Typical examples of the density patterns of the stable zero-vorticity
onsite-centered and offsite-centered LQDs are shown in Fig. \ref{2}, which
correspond to points \textquotedblleft a, b, c, and d\textquotedblright\ in
the stability areas in Figs. \ref{1} (a,b). The parameters are $(N,V_{0}%
)=(3,-2)$, $(22,-2)$, $(10,-2)$, and $(50,-2)$, respectively. Note that the
LQDs in Figs. \ref{2}(a2,a3) and (b2,b3) are selected from the bistable areas,
which allows the coexistence of different numbers of sites with equal norms
and lattice modulation depths, see the points \textquotedblleft
b\textquotedblright\ and \textquotedblleft d\textquotedblright\ in Figs.
\ref{1}(a,b), respectively. It is worth noting that there are some other LQDs
solutions, which are not symmetric to the origin of the coordinate are found
outside the color stripe areas, i.e., the blank region. Typical examples of
these types of LQDs are shown in Fig. \ref{3}, which correspond to points
\textquotedblleft e and f\textquotedblright\ in the blank areas in Figs.
\ref{1}(a, b). As seen from the evolution figure [Figs. \ref{3}(a2, b2)],
these types of LQDs are also stable. Obviously, these types of LQDs cannot be
characterized by the expansion law in Fig. \ref{0}; these types will not be
discussed in detail in the current paper.

Fig. \ref{1}(c) displays the dependence of the chemical potential $\mu$ and
the total norm $N$, for stable onsite-centered and offsite-centered LQDs,
here, we fixed $V_{0}=-2$. The results indicate that only the $\mu\left(
N\right)  $ curves for $n=1$ (onsite-centered LQDs) satisfy $d\mu/dN<0$ when
$N<1.8$, which is well known as the Vakhitov-Kolokolov (VK) criterion (the
necessary stability condition for a soliton in the attractive background).
However, when $N>1.8$, the $\mu(N)$ curves for the LQDs with all the numbers
of sites satisfy $d\mu/dN>0$, which seems to violate the VK criterion. The
similar results were previously reported in 1D model and the simulations have
demonstrated that both soliton branches are stable in the CQ-nonlinear channel
\cite{PRE71_DisobeyVK}. While, in 2D model, which combines a checkerboard
potential, alias the Kronig-Penney lattice, with the self-focusing cubic and
self-defocusing quintic nonlinear terms, the branchs of each bistable solitons
are stable against arbitrary perturbations \cite{PRE74_DisobeyVK}. Next, we
will provide the analysis to explain why the $\mu(N)$ curves for these LQDs
can violate the VK criterion.

According to Eq. (\ref{Model2}), the stationary solution in the vicinity of
$\phi_{p}$ obeys the stationary GP equation as follows:%
\begin{equation}
\mu\phi_{p}=\frac{V_{0}}{2}\phi_{p}+\phi_{p}|\phi_{p}|^{2}\text{ln}|\phi
_{p}|^{2}. \label{A2}%
\end{equation}
Here, we have applied the TF approximation in Eq. (\ref{A2}). According to Eq.
(\ref{totalarea}), the total $\phi_{p}$ can be expressed as $|\phi_{p}%
|^{2}\approx N/\mathcal{A}$. Therefore, the chemical potential $\mu$ obeys%
\begin{equation}
\mu=\frac{V_{0}}{2}+\frac{N}{\mathcal{A}}\text{ln}\frac{N}{\mathcal{A}}.
\label{A3}%
\end{equation}
Hence,
\begin{equation}
{\frac{d\mu}{dN}}={\frac{1}{\mathcal{A}}}\left[  \ln\left(  {\frac
{N}{\mathcal{A}}}\right)  +1\right]  , \label{A4}%
\end{equation}

From Eq. (\ref{A4}), when $\ln\left(  N/\mathcal{A}\right)  +1>0$, i.e.,
$N>\mathcal{A}/e$, we can obtain%
\begin{equation}
\frac{d\mu}{dN}>0. \label{A5}%
\end{equation}
So, above analysis demonstrate the $\mu(N)$ function for these LQDs can
violate the VK criterion. In our simulations, we choose the lattice constant
$D=5$, so the area per lattice site $s=\pi(D/4)^{2}\approx4.91$, hence,
$\mathcal{A}=1\ast s=4.91$, yields, $N_{cr\_th}=\mathcal{A}/e=1.806$. From
Fig. \ref{1}(c), it can be seen that the numerical results ($N=1.8$) are
basically in agreement with the theoretical prediction ($N_{cr\_th}%
\approx1.806$).

\subsection{Vortex LQDs with $S=1$}

\begin{figure}[ptb]
\centering
\includegraphics[scale=0.3]{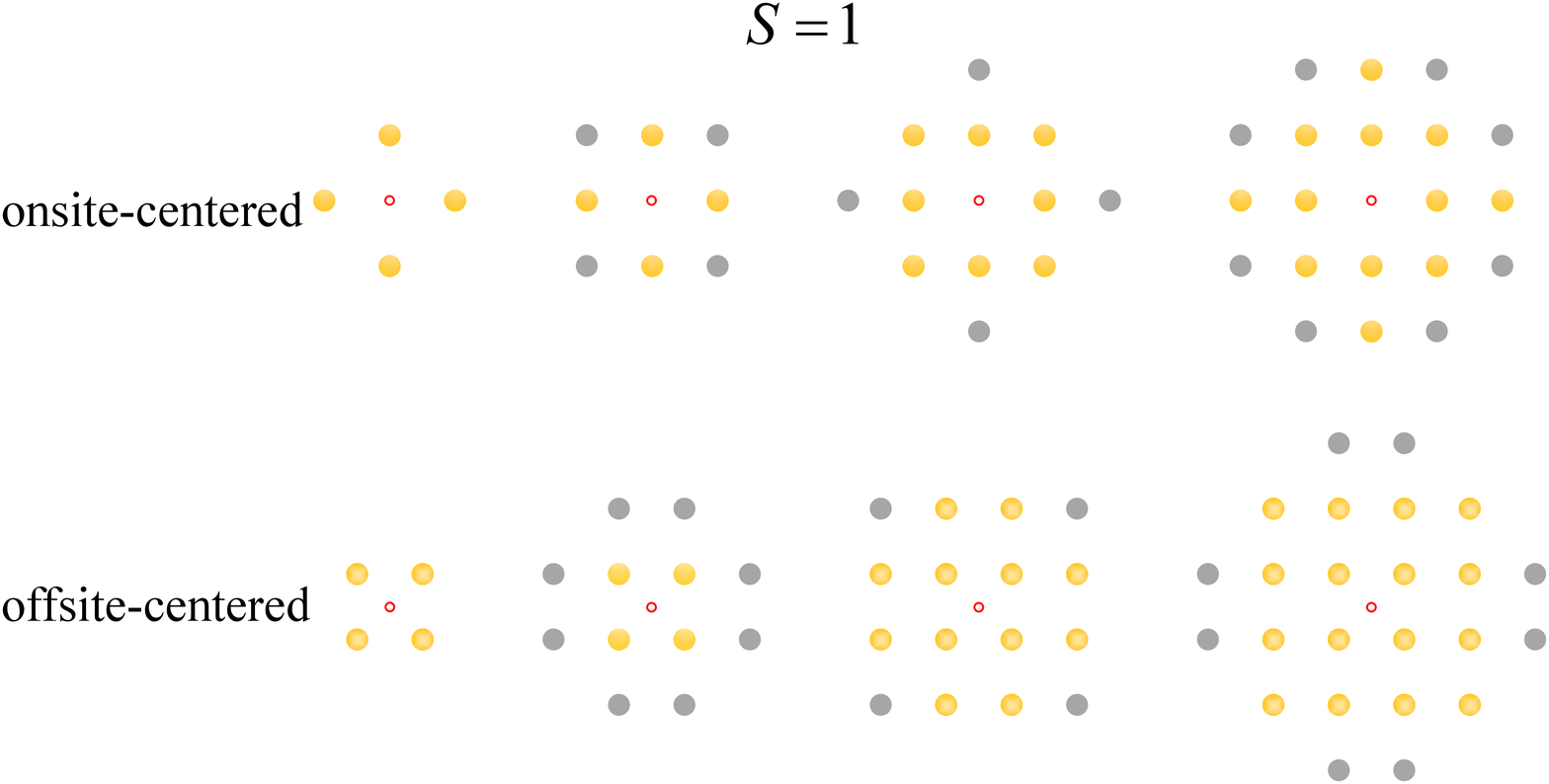}\caption{The upper row is the sketch map
of the growth pattern of onstie-centered LQDs with $S=1$ versus $n$. The lower
row is the same sketch map for offsite-centered LQDs. The red circles
represent the centers of the LQDs.}%
\label{S1}%
\end{figure}

Whether the stable vortex LQDs can be found in current systems is a nontrivial
issue. Generally, a vortical mode must have a density pivot at its center.
Hence, for the onsite-centered vortex LQDs, their center lattice site will be
occupied by the pivot. For offsite-center vortex LQDs, because the density of
their center is already zero, it is not necessary to provide a lattice site to
the density pivot. Fig. \ref{S1} shows the sketch maps of the growth patterns
of the first 4 steps for\ the two types of vortex LQDs. For the
onsite-centered vortex LQDs, because the central lattice site is already
occupied by the density pivot, the $1^{\mathrm{st}}$ step of its pattern
occupies the 4 closest lattice sites (i.e., $n=4$) at the origin of the
coordinates. Then, it expands to $n=8$, $12$ and $20$ in the next 3 steps.
This process is similar to the zero-vorticity case starting from the
$2^{\mathrm{nd}}$ step with the central lattice site being removed. For the
offsite-centered vortex LQDs, because the density is zero at the center, the
expansion law is the same as that in the zero-vorticity case in Fig. \ref{0}.
Thus, we can see that the offsite-centered LQDs with zero-vorticity and vortex
LQDs with $S=1$ are heterogeneous, which means that they occupy the same
lattices with the same density pattern, but have different amplitude and phase
distribution. It is also interesting to find that the numbers of sites are
even for both the \ onsite-centerred and offsite-centered types. In the
following discussion, we will continue to use the number of sites to
characterize the vortex LQDs.

\begin{figure}[ptb]
\centering
\includegraphics[scale=0.8]{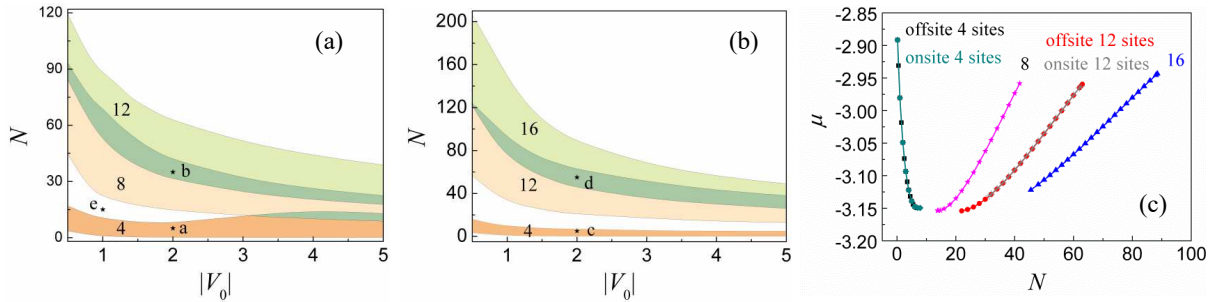} \caption{(a, b) Stability area of the
three minimum isotropic vortex LQDs with $S=1$, which correspond to $4$, $8$,
and $12$ sites for the onsite-centered LQDs and $4$, $12$, and $16$ sites for
the offsite-centered LQDs in the plane of ($N$, $\left\vert V_{0}\right\vert
$). The bistability states occur in the light green areas. (c) Chemical
potential $\mu$ versus the total norm $N$, for the stable onsite-centered and
offsite-centered LQDs. Here, we fixed $V_{0}=-2$.}%
\label{4}%
\end{figure}

\begin{figure}[ptb]
\centering
{\includegraphics[width=0.8\columnwidth=1]{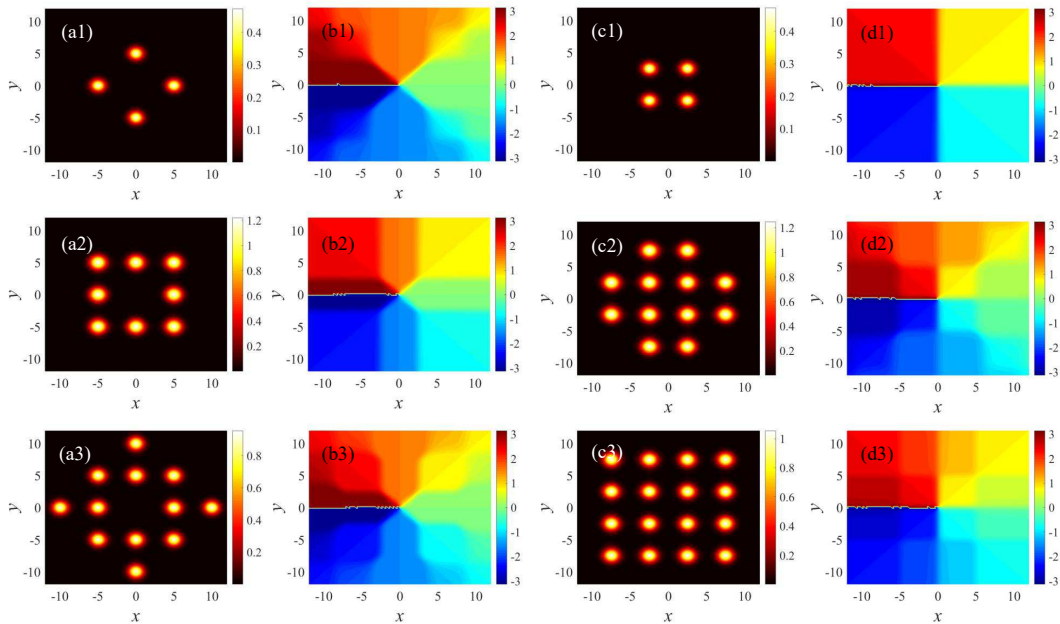}}\caption{(Color online)
Typical examples of the density patterns of the stable onsite-centered (a1-a3)
and offsite-centered (c1-c3) vortex LQDs with $S=1$, which correspond to
points \textquotedblleft a, b, c, and d\textquotedblright\ in the stability
areas in Fig. \ref{4}. The parameters are $(N,V_{0})=(5,-2)$, $(35,-2)$,
$(5,-2)$, and $(55,-2)$, respectively. (b1-b3) and (d1-d3) show the
corresponding phase patterns of (a1-a3) and (c1-c3), respectively. Here, the
LQDs in (a2,a3) and (c2,c3) are selected from the bistable areas [see the
points \textquotedblleft b\textquotedblright\ and \textquotedblleft
d\textquotedblright\ in Figs. \ref{4}(a,b) respectively].}%
\label{5}%
\end{figure}

\begin{figure}[ptb]
\centering
{\includegraphics[width=0.8\columnwidth=1]{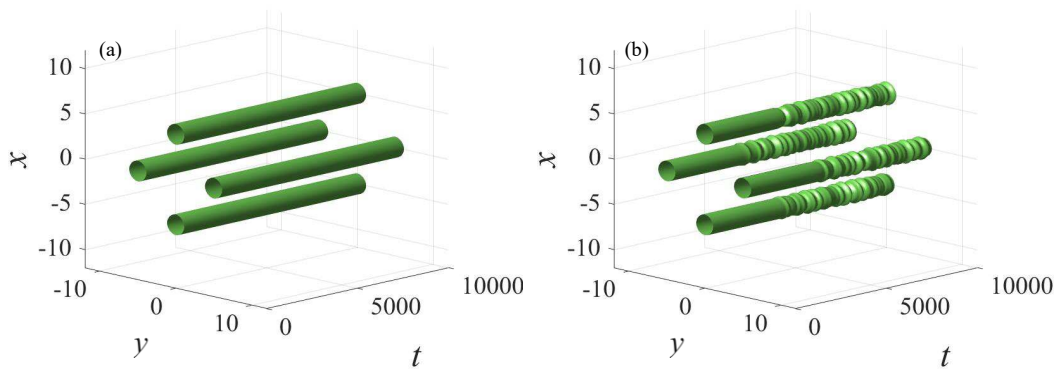}}\caption{(Color online)
(a) Direct simulations of the evolution of stable onsite-centered vortex LQDs
with $S=1$, which correspond to point \textquotedblleft a\textquotedblright%
\ in the stability areas in Fig. \ref{4}(a). (b) Typical examples of unstable
LQDs in the white region, which correspond to point \textquotedblleft
e\textquotedblright\ in the stability areas in Fig. \ref{4}(a), and parameters
are $(N,V_{0})=(15,-1)$.}%
\label{6}%
\end{figure}

Numerical results show that stable vortex LQDs with $S=1$ exist in our
systems. Similar to LQDs with zero-vorticity, we plot their stability areas
with different colors in the ($N,|V_{0}|$) plane in Figs. \ref{4}(a, b).
Similar to the discussion above regarding zero-vorticity, these color stripes
concentrate with increasing modulation depth $|V_{0}|$. Hence, the overlaps
between them, which are labeled by the light green areas, also appear. Typical
examples of the density patterns of the stable onsite-centered (a1-a3) and
offsite-centered (c1-c3) vortex LQDs with $S=1$ are shown in Fig. \ref{5},
which correspond to points \textquotedblleft a, b, c, and d\textquotedblright%
\ in the stability areas in Fig. \ref{4}. The LQDs in (a2,a3) and (c2,c3) are
selected from the bistable areas, which allows the coexistence of different
numbers of sites with equal norms and lattice modulation depths, see the
points \textquotedblleft b\textquotedblright\ and \textquotedblleft
d\textquotedblright\ in Figs. \ref{4}(a,b) respectively. Figs. \ref{5}(b1-b3,
d1-d3) show the corresponding phase patterns. Here, we fixed the lattice
modulation depth $V_{0}=-2$. Outside the color areas, unstable vortex LQDs may
be found. Typical examples of the evolution of the stable and unstable vortex
LQDs are shown in Fig. \ref{6}, which correspond to points \textquotedblleft
a\textquotedblright\ and \textquotedblleft e\textquotedblright,
respectively,\ in the stability areas in Fig. \ref{4}(a).

Fig. \ref{4}(c) displays the dependence of the chemical potential $\mu$ and
the total norm $N$, for stable onsite-centered and offsite-centered vortex
LQDs. Here we fixed $V_{0}=-2$. It is worth noting that the two types of LQDs
with the same numbers of sites will have equal chemical potentials at the same
norm $N$ [see Fig. \ref{4}(c), the 4-site and 12-site onsite-centered and
offsite-centered vortex LQDs, respectively]. This phenomenon indicates that at
the same norm $N$, these two types of LQDs with the same numbers of sites are
degenerate states. This phenomenon is different from the LQDs with
zero-vorticity. Next, as with the zero-vorticity LQDs, the dependencies
between the chemical potential $\mu$ and the total norm $N$, does not
necessarily satisfy the VK criterion.

\section{conclusion}

In this paper, we studied two-dimensional (2D) lattice quantum droplets (LQDs)
trapped in 2D optical lattices. First, we demonstrated the possibility of
creating stable zero-vorticity and vortex LQDs and then investigated the
influence of the optical lattice potential on the LQDs. We found two types of
stable LQDs: onsite-centered and offsite-centered LQDs. Furthermore, the
stability areas of the two types of zero-vorticity LQDs and vortex QDs with
$S=1$ are given. Bistability characteristics are shown in a stable region
diagram, which allows the coexistence of different numbers of sites with equal
norms and lattice modulation depths. Then, we found that some other
zero-vorticity LQDs are stable in the blank region. Different from the
zero-vorticity LQDs, the $\mu-N$ relationship shows that the two types of
vortex LQDs with the same number of sites are degenerated. An important point
is that for stable LQDs with a fixed number of sites, the dependencies between
the chemical potential $\mu$ and the total norm $N$, can violate the
Vakhitov-Kolokolov (VK) criterion.

The present analysis can be extended in some directions. First, the current
study is based only on square lattices, and we can extend it to other lattice
structures, such as triangle lattices or graphene structures. The latter
option may be able to relate the LQDs to topological phenomena. Further, for
the vortices, it is worth considering whether hidden vortices and anisotropic
vortices can be supported by the lattice. Finally, a challenging option is to
seek stable vortex LQDs in the 3D configuration.

\section{Acknowledgments}

This work was supported by the National Natural Science Foundation of china
(NNSFC) through grant nos. 11905032, 11874112, the Key Research Projects of
General Colleges in Guangdong Province through grant no. 2019KZDXM001, the
Foundation for Distinguished Young Talents in Higher Education of Guangdong
through grant No. 2018KQNCX279, and the Special Funds for the Cultivation of
Guangdong College Students Scientific and Technological Innovation, No. xsjj202005zra01.

\section{references}

\end{document}